\documentclass{article}
\def\ben{\begin{equation}}
\def\een{\end{equation}}

\def\bea{\begin{eqnarray}}
\def\eea{\end{eqnarray}}
\def\p{\partial}

\def\br{{\bf r}}

\def\bn{{\bf n}}

\usepackage[dvips]{graphicx}
\begin{document}

\hfuzz=100pt
\title{The evolution of pebble size and shape in space and time}
\author{G. Domokos \\
{\em Department of Mechanics, Materials, and Structures }\\
{\em Budapest University of Technology and Economics}\\
{\em M\"uegyetem rkp.3 }\\
{\em Budapest, 1111 Hungary} \\
{\em and}  \\
G. W. Gibbons  \\
{\em D.A.M.T.P.}\\
{\em  Cambridge University}\\
{\em Wilberforce Road, Cambridge CB3 0WA, U.K.}\\}

\date{\today}       
 \maketitle

\begin{abstract}
We propose a mathematical model which suggests that the two main geological
observations about shingle beaches, i.e. the emergence of predominant
pebble size ratios and strong segregation by size are interrelated. Our
model is a based on a system of ODEs called the \em box equations, \rm describing
the evolution of pebble ratios. We derive these ODEs as a heuristic approximation of
Bloore's PDE describing collisional abrasion and verify them by simple experiments and
by direct simulation of the PDE.
While representing a radical
simplification of the latter, our system admits the inclusion of additional
terms related to frictional abrasion. We show that
nontrivial attractors (corresponding to predominant pebble size ratios)
only exist in the presence of friction. By interpreting our
equations as a Markov process, we illustrate by direct simulation that these
attractors may only be stabilized by the ongoing segregation process.
\end{abstract}


\section{Introduction}
The shape of pebbles has been a matter of discussion
since at least the time of Aristotle \cite{Krynine1}. 
In general, the central question is whether particular
pebble shapes emerge from the abrasion and transport processes.
Aristotle himself  claimed  that spherical shapes dominate.
However, as Aristotle also observed,  abrasion is a complex 
interaction between the abraded
particle and the abrading environment  
represented by 'other objects' (i.e. other pebbles)
where not only local properties (e.g. curvatures)
but also semi-global effects due to particle shape as 
well as global effects due to particle transport
play important role. In this process pebbles 
mutually abrade each other, defining the time evolution both for the
abraded pebble and the abrading environment represented by
other particles subject to particle transport.  
In the simplest approach, one neglects the latter effects
and regards the abrasion of a single pebble in a  constant environment. 
Aristotle's model for individual abrasion  may be translated as
\ben \label{A1}
v = f(R) \,,
\een
where $v$ is the speed  with which the pebble's surface moves along the inward normal, 
$R$ is the radial distance from the center of gravity of the pebble being abraded, $f$ is a monotonically
increasing function of $R$ only,  and in particular, $f$ is  independent
 of time. Note that
 (\ref{A1}) is a partial integro-differential equation,
since the location of the center of gravity is 
determined by time-dependent integrals.
 The modern , PDE-based,
theory of individual abrasion appears to start with 
the pioneering 
work, both experimental
and theoretical,  of Lord Rayleigh \footnote{Son of the Nobel Prize winner
Lord Rayleigh}  \cite{Rayleigh1,Rayleigh2,Rayleigh3}.
For some earlier work see \cite{Palmer,Dobkins}. 
Rayleigh  mainly considered axisymmetric pebbles and 
he observed that the ultimate shape was not necessarily, indeed seldom,
spherical.   
He found that  some pebble's shapes are far from ellipsoidal, being much more
discoid in shape. He asserted that abrasion cannot be a simple
function of the Gauss (or {\it specific}) curvature
$K$ \index{specific curvature}.

On the mathematical side, Firey \cite{Firey} initiated a 
rigorous study by adopting a PDE  model rejected by Rayleigh
in which the shape evolved according to what  Andrews \cite{Andrews} calls
the {\sl flow by the Gauss curvature} that is, he studied the PDE 
\ben \label{Firey1}
v = cK \,,
\een
where $c$ is a constant and proved that all convex shapes ultimately converge to the sphere
under the action of (\ref{Firey1}). Note that the word \em flow \rm is being
used in the sense in dynamical systems theory and it should not be
confused with physical fluid flow.
Later, this proof was substantially amplified by Andrews (\cite{Andrews}). 
Recently, Durian \cite{Durian}
investigated the statistical distribution of Gaussian curvature on pebble shapes.

The physical assumption underlying Firey's model is that 
the abraded particle (pebble) undergoes a series
of small collisions with a very large, smooth abrader, and this might be the case 
when pebbles are carried by
a fast river and collide repeatedly with the riverbed, a process called bedload transport.
This  concept of \em collisional flows \rm was substantially generalized by 
Bloore  who  studied abrasion by
spherical abraders of radius $r$ (Firey's model corresponds to $r \to \infty$).
With brilliant intuition Bloore 
\cite{Bloore} arrived at the PDE
\ben \label{Bloore1}
v = 1+2bH+cK \,,
\een
where $b$ and $c$ represent, as Bloore describes, the average, or 'effective' values for $r$ and $r^2$, respectively and $H$ is the Mean Curvature.
Bloore also showed that a spherical shape of radius $R$ abraded by an abrader of radius $r$ is stable under the action of (\ref{Bloore1}) if
\ben \label{Bloore2}
\frac{R}{r} < 3\,.
\een
Apparently, (\ref{Bloore1}) consists of three terms: the constant first term which corresponds to
the so-called Eikonal equation, the Mean Curvature term $2bH$ and the Gaussian term $cK$. The latter was
studied by Firey, while for the 
Mean Curvature Flow Huisken \cite{Huisken} showed that it also converges
to the sphere. In the case of the Eikonal we know that the 
\em outward \rm flow (which can be imagined also
as time-reversed abrasion) converges to the sphere,
so, simply put, the question is whether the three terms can 'balance' i.e. homothetic solutions
can exist.

The rigorous foundation of the collisional model requires a 
probabilistic approach as it was already 
conjectured by Hilbert \cite{Hilbert}. 
Based on results
from Stochastic and Integral  Geometry \cite{SchneiderWeil} one can not only justify
Bloore's equations but also identify the physical meaning of the coefficients
 $b,c$ in (\ref{Bloore1}) \cite{VarkonyiDomokos}:
  
\ben \label{ibc}
b= \frac{M}{4 \pi }, \qquad  c = \frac{A}{4 \pi} \,,
\een  
where $M$ is the integrated  mean curvature and 
$A$ the area of the {\sl abrading}  particles, assumed convex and identical. In case of spherical
abraders this is identical with Bloore's interpretation of the coefficients.
By {\it Minkowski's inequality}   for  convex bodies   \cite{Minkowski} the quantities
$b$ and $c$ may not be given arbitrarily, rather, based on (\ref{ibc}) they must satisfy
\ben \label{VD1}
b^2 \ge c \,.
\een
The main question is now whether  collisional flows defined in (\ref{Bloore1}), (\ref{ibc}) can explain
the observation of spherical and non-spherical pebbles shapes, 
i.e. whether (\ref{Bloore1}), (\ref{ibc})
admits homothetic solutions and what they look like.

Independently of  the mathematical work which we have 
briefly reviewed, there has been
considerable geological interest in pebble shape which partly
focused on individual pebble shapes, partly on global understanding
of pebble transport and spatial distribution on large shingle beaches.
An extensive list of references about individual pebble shape up to 1970  
is contained
in \cite{Dobkins}.

One of the most remarkable accounts of pebble shape and 
distribution is provided
by Carr \cite{Carr} based on the measurement of approximately a hundred thousand pebbles on Chesil Beach, Dorset, England. 
In summarizing his results, Carr first
notes that \em a pebble appears to retain its proportions irrespective of size
and angularity. \rm  Apparently, the same phenomenon was described by Landon  \cite{Landon} claiming that coastal pebbles
evolve towards what he calls \em equilibrium shapes.\rm  The subject was taken up  in the 1990's in {\it Nature}
\cite{Ashcroft,Lorang,Yazawa}   
following a note by Wald \cite{Wald} claiming that
pebbles with three planes of symmetry typically had
the form of flattened ovoids in proportion 7:6:3. 
In the absence of pebble transport, these observations would imply the existence of stable (attracting) homothetic (self-similar) solutions of the governing equations.
As we will point out, in the absence of transport
such solutions do not exist. 

However, Carr also provides mean values and sample
variations for maximal pebble size along lines orthogonal to the beach.
These plots reveal pronounced segregation by maximal size, i.e. on shingle
beaches pebbles of roughly similar maximal sizes appear to be spatially
close to each other. This phenomenon, also confirmed by  \cite{Bird},
\cite{Gleason},\cite{Hansom}, \cite{Kuenen_Grading},\cite{Landon}, \cite{Neate}
 is closely related to the global
transport of pebbles by waves \cite{Lewis}, \cite{Carr1}. Grading  not
only by size but also by
shape has been also reported, the most notable description is due to Bluck \cite{Bluck}, reporting on what he calls \em equilibrium states \rm
of pebble collections corresponding to stationary shape
distributions (cf also \cite{Landon}, \cite{Orford}, \cite{Williams}).

Apparently, the distribution of pebble shapes and size are
controlled by two fundamental geological processes: particle abrasion
and particle transport.  Which of these processes
dominates may depend on the
geological location and also on time, however, geologists appear to agree
that the role of transport may not be neglected. Another general observation is
that this process approaches equilibria in the form of stationary
pebble shape distributions with dominant peaks.
A detailed account of the interaction
of abrasion and transport is given by Landon \cite{Landon} investigating the beaches on the west shore of Lake Michigan. He attributes shape variation to  a mixture
of abrasion and transport. Kuenen \cite{Kuenen_Surf} discusses Landon's observations, however disagrees with the conclusions and attributes shape variation primarily to transport. Carr \cite{Carr} observes dominant shape ratios
emerging as a result of abrasion and size grading while Bluck \cite{Bluck}
describes beaches in South Wales where equilibrium distributions of shape
are reached primarily by transport and abrasion plays a minor role.

These observations call for a mathematical model where both abrasion
and transport are included and one may study their complex interaction
in the entire range from pure abrasion to pure segregation.
Our goal in this paper is to make one step towards such a general model which is based on the classical theory of particle abrasion but also admits the inclusion
of global transport. 
We derive a system  of representative  ODE's called the {\it box  equations}  
which  provide a huge simplification of Bloore's
PDE's but which nevertheless  allow us 
to study the spatio-temporal process. In this framework we can show that 
in accord with the observations of Landon, Carr and Bluck, in the absence
of global transport and segregation, stationary distributions
with sharp peaks (centered around homothetic solutions of the
deterministic equations) \em may not emerge \rm in a stable manner.
After verifying our analytical results both by direct simulation of Bloore's PDE and by simple experiments 
we also show that abrasion and size grading alone (most clearly stated by Carr) might be sufficient to produce persistent dominant basic proportions. 
Strong shape grading (as described by Bluck) enhances this process even further.
Our model creates a framework where both extreme cases (pure abrasion in the absence of transport and pure transport in the absence of abrasion),
as well as their combination may be studied.

\section{The Collisional Box Equations} \label{s_box}

\begin{figure}[!ht]
\begin{center}
\includegraphics[width=100 mm]{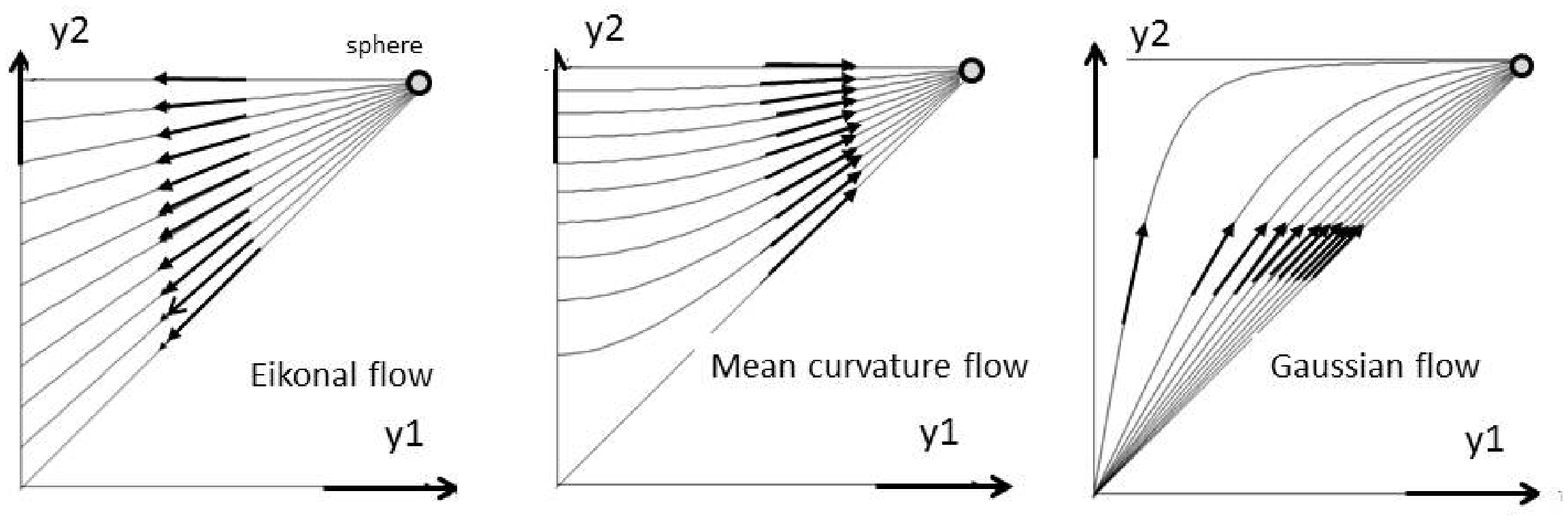}
\end{center}
\caption{2D component flows: Eikonal, Mean Curvature and Gaussian flows}\label{fig:components}
\end{figure}

Motivated partly by the above described geological observations,
instead of the numerical study of (\ref{Bloore1}) we propose  
 representing   shapes by orthogonal
bounding boxes (defined by 6 orthogonal bounding planes)
with sizes $2u_1\leq2u_2\leq2u_3$, so  that the
  attrition
speed $v$ is replaced by $du_i/dt=\dot{u_i}$, the local quantities $K,H$ 
we obtain from the ellipsoid which fits   inside the box:
\ben
K_i=\frac{u_i^2}{u_j^2u_k^2},\quad H_i=\frac{1}{2}\left(\frac{u_i}{u_j^2}+\frac{u_i}{u_k^2}\right), \quad i\not=j\not=k\,.
\een
 Since we are interested in the
existence of self-similar solutions,
we introduce the box ratios $y_1=u_1/u_3, y_2=u_2/u_3$, we denote the origin $(y_1,y_2)=(0,0)$ by $O$
the sphere at $(y_1,y_2)=(1,1)$ by $S$
and to simplify notation we write $y_3=u_3$. This approach has the advantage
that most of the available geological data is expressed in terms of these variables, in fact, the representation in
the $(y_1,y_2)$ plane is often referred to as the \em Zingg Triangle \rm
\cite{Zingg}. Now the  PDE  
(\ref{Bloore1}) is reduced to the system of O.D.E,'s
 in the $(y_1,y_2,y_3)$ system:

\bea 
\label{yy_box3D}
\dot y_i & = & F_i(y_1,y_2,y_3,b,c) =\frac{F_i^E}{y_3} +2 b \frac{F_i^M}{y_3^2}  + c \frac{F^G_i}{y_3^3} \\
\label{y3_box3D}
-\dot y_3 & = & F_3(y_1,y_2,y_3,b,c)= 1 + \frac{b}{y_3} \frac{y_1^2+y_2^2}{y_1^2y_2^2}+\frac{c}{y_3^2}\frac{1}{y_1^2y_2^2}\,,
\eea

where
\ben \label{components}
F_i^E= y_i-1, \hspace{0.5cm}F_i^M=\frac{1-y_i^2}{2y_i}\hspace{0.5cm}F_i^G=\frac{1-y_i^3}{y_iy_j^2}, \hspace{0.5cm}i,j=1,2; i\not=j \,.
\een
As  can be  seen  from (\ref{yy_box3D})-(\ref{y3_box3D}), the full
flow is three dimensional, however, the component flows (Eikonal,
Mean curvature, Gaussian) have autonomous planar components (illustrated in figure
\ref{fig:components}):
\ben \label{component_flows}
 y'^E_{2} = \frac{F_2^E}{F_1^E}=\frac{y_2-1}{y_1-1}, \quad y'^M_{2}  = 
 \frac{F_2^M}{F_1^E} =\frac{y_1(y_2^2-1)}{y_2(y_1^2-1)}, \quad
 y'^G_{2}  =\frac{F_2^G}{F_1^G}=  \frac{y_2(y_2^3-1)}{y_1(y_1^3-1)}\,,
\een 
where $()'=d/dy_1$.
By introducing the vector notation $\mathbf{y}=[y_1,y_2,y_3]^T, \mathbf{F}=[F_1,F_2,F_3]^T$, 
(\ref{yy_box3D})-(\ref{y3_box3D}) can be rewritten as
\ben
\label{i_individual_box3D}
\dot{\mathbf{y}}=\mathbf{F}(\mathbf{y},b,c)\,.
\een
Our main concern is the detection of fixed points and self-similar (homothetic) solutions.
According to (\ref{y3_box3D}) $\dot y_3<0$, so (\ref{yy_box3D})-(\ref{y3_box3D}) does not have
fixed points. Homothetic solutions satisfying
\ben \label{homo}
\dot y_1(t)\equiv\dot y_2(t)\equiv 0
\een
may exist, in fact spheres at $S$ are a homothetic  solution of (\ref{yy_box3D})-(\ref{y3_box3D}).
Apart from  spheres, (\ref{yy_box3D})-(\ref{y3_box3D})
will  only  admit  homothetic solutions if (\ref{yy_box3D}) is
autonomous and has fixed points; we explore this case in subsection \ref{s_selfdual}. If
(\ref{yy_box3D}) is not autonomous, then homothetic solutions do not exist, 
nevertheless
one still may look for solutions of 
\ben \label{loiter}
\dot y_1(t_0)=\dot y_2(t_0)=0
\een
as 'temporary' homothetic solutions, we explore this in subsection \ref{s_stat}. Whether permanent
or temporary, the existence of homothetic solutions requires that the $(y_1,y_2)$
projection of the flow should be balanced and this observation admits the formulation
of a \em necessary \rm condition. By considering that the flow is a linear combination
of three components (\ref{component_flows}) two of which converge to sphere, we can observe that
if either (\ref{homo}) or (\ref{loiter}) are to be satisfied then
the slope of the Eikonal should be 'bracketed' by the two other slopes:
\ben \label{neccond_collision}
(y'^E_{2}- y'^G_{2})(y'^E_{2}- y'^M_{2})<0\,.
\een
Based on (\ref{component_flows}) we can readily verify that in the interior $y_1<y_2<1$ of the Zingg
triangle this is indeed the case since
\ben \label{bracket1}
y'^M_{2} < y'^E_{2} <y'^G_{2} <\frac{y_2}{y_1}\,,
\een
i.e. the necessary condition (\ref{neccond_collision}) is always fulfilled, cf. Figure \ref{fig:neccond}/a.

Box flows represent only a heuristic approximation of the PDE
 and we may use 
several alternative definitions
of the bounding box when measuring the box dimensions. Selection of
special approximations based on extremal
properties has been studied and a famous example
is the unique John ellipsoid with maximal volume \cite{John}.
Box equations are the basic tool enabling us to follow Aristotle's
ideas and to include semi-global and global effects in
our model. On the other hand, box equations are Aristotelean also in the local sense since in the
box approximations of the Bloore flows (\ref{Bloore1})
attrition speeds $du_i/dt=\dot{u_i}$ are a monotonically growing function of the distance $u_i$.
When deriving the box ODE's   from the  PDE
 (\ref{Bloore1}), the abrading environment (represented
by the coefficients $b,c$)
can also be approximated by its box dimensions $v_1\leq v_2\leq v_3$
and box ratios $z_1=v_1/v_3,z_2=v_2/v_3$ and maximal box size $z_3=v_3$. 
According to (\ref{ibc}) the coefficients $b,c$ depend on the surface integrals $M,A$, and the latter
we derive directly  from the abrading orthogonal box, with ratios $z_i$:
\ben \label{integ}
M  = 2\pi z_3(z_1+z_2+1) \,, \quad
A  = 8z_3^2(z_1z_2+z_1+z_2)\,.
\een
The same quantities can be expressed for the unit cube as 
$M_1 = 6\pi, \quad  A_1  =  24$,
so we have
\ben \label{coeffs}
b(\mathbf{z})= \frac{M}{M_1}  =z_3\frac{z_1+z_2+1}{3}= z_3f^b_z\,,\quad
c(\mathbf{z}) = \frac{A}{A_1}=z_3^2\frac{z_1+z_2+z_1z_2}{3}= z_3^2f^c_z\,.
\een
We remark that (\ref{VD1}) remains valid also in the box approximations. While the PDE model (\ref{Bloore1})
is describing the collisional abrasion of an \em individual \rm pebble in the abrading environment represented by the coefficients $b,c$, 
the box equations suggest a natural generalization which tracks the \em mutual \rm abrasion of two pebbles $\mathbf{y},\mathbf{z}$,
each of which represents the abrading environment for the other. (Alternatively, one may think of two pebble \em types \rm abrading each other. We will return to the proper statistical interpretation of these equations in section \ref{s_stochastic})
Based on
(\ref{i_individual_box3D}) and (\ref{coeffs}) we obtain the simultaneous ODEs:
\bea
\label{y_box3D}
\dot{\mathbf{y}}&=&\mathbf{F}(\mathbf{y},b({\mathbf{z}}),c({\mathbf{z}}))=\bar \mathbf{F}(\mathbf{y},\mathbf{z})\\
\label{z_box3D}
\dot{\mathbf{z}}&=&\mathbf{F}(\mathbf{z},b({\mathbf{y}}),c({\mathbf{y}}))=\bar \mathbf{F}(\mathbf{z},\mathbf{y})
\eea
describing the (approximate) evolution of box ratios under a purely collisional abrasion process.
At first sight, the mutual abrasion of two pebbles may not seem to be a realistic model for the shape evolution of vast pebble collections
on shingle beaches, nevertheless, we will show that equations (\ref{y_box3D})-(\ref{z_box3D}) open up the possibility
 of  describing  the global interaction 
between abraded pebble and the abrading environment, including segregation.
Another, important
feature of the box equations is that they can be easily extended to model abrasion beyond collisional
processes. First we look at two special collisional processes.

\subsection{The stationary case}\label{s_stat}

First we study the case where pebble $\mathbf{z}$ is infinitely harder then pebble $\mathbf{y}$,
i.e. we have $\mathbf{z}(t)\equiv\mathbf{z}(0)$ so (\ref{y_box3D})-(\ref{z_box3D}) is reduced to the 3D system:
\ben
\label{steady}
\dot \textbf{y} =  \bar \textbf{F}(\textbf{y},\textbf{z}(0))\,,
\een
which we call the \it steady state flow. \rm Since here we have constant coefficients $b,c$, this
case is directly comparable to the  PDE  (\ref{Bloore1}).
The infinitely hard pebble $\mathbf{z}$ could represent either the average
value of a very hard pebble population in which the relatively soft pebble $\mathbf{y}$ is being abraded
or, alternatively, the average value of a pebble population which
remains invariant not because every pebble remains invariant but because
it has a constant source and a constant sink. In the latter interpretation $\mathbf{y}$
is also a member of this population. We are interested in the geometric properties
of these flows. While the explicit formulae in (\ref{y_box3D})-(\ref{z_box3D}) admit
a rather straightforward, rigourous mathematical study, here we merely state the main geometric facts as observations.

\begin{figure}[!ht]
\begin{center}
\includegraphics[width=120 mm]{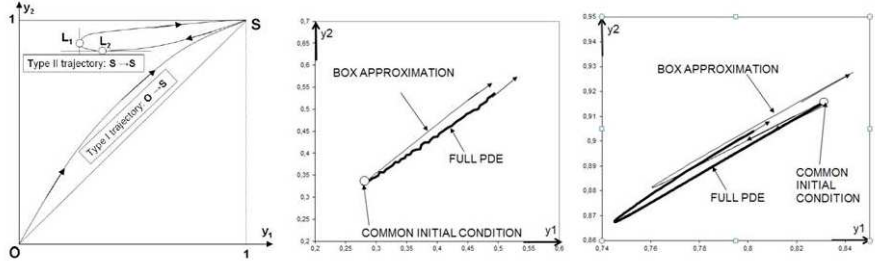}
\end{center}
\caption{\small{Geometry of the steady state flows. a) Type I and Type II trajectories in the Zingg triangle. Turning points in $y_1,y_2$ are marked with $L_1,L_2$, respectively. b) Simulation of Type I trajectory: ellipsoid with axis ratios $y_1=0.278,y_2=0.329, y_3=1.00$ evolved under the PDE (\ref{Bloore1}) and under the steady state box flows (\ref{steady}) in the presence of constant spherical abraders with radius $r=0.5$ ($z_1=z_2=1, z_3=0.5, b=0.5, c=0.25$) c) Simulation of Type II trajectory: ellipsoid with axis ratios $y_1=0.83,y_2=0.91, y_3=1.00$ evolved under the PDE (\ref{Bloore1}) and under the steady state box flows (\ref{steady}) in the presence of constant spherical abraders with radius $r=0.1$}} \label{fig:PDE_LOOP}
\end{figure}

As stated in section \ref{s_box}, the flow (\ref{steady}) can not have any 
genuine fixed points, and the only homothetic solution is the sphere $S$. Nevertheless,
to explore the geometry of the flow we can still ask whether 'temporary' homothetic solutions,
satisfying (\ref{loiter}) may exist thus we seek the simultaneous solution of the quadratic
equations
\ben 
\label{quadratic}
y_3 ^2 F_i^E +2 b y_3 F_i^M + c F^G_i = 0\, \quad i=1,2\,. 
\een
Two out of four roots of (\ref{quadratic}) are strictly positive, two strictly negative. The former
define two surfaces $\lambda_i(y_1,y_2,b,c)$, $(i=1,2)$, where
$b,c$ are substituted from (\ref{coeffs}).\em Above \rm these surfaces 
($y_3>\lambda_i$) the flow moves
away from $S$, towards $O$, below the surfaces ($y_3<\lambda_i$) the opposite happens, between the surfaces
the derivatives $\dot y_1, \dot y_2$ have opposite signs. Points along the intersection lines defined by
 $\lambda_1=\lambda_2$ satisfy (\ref{loiter}), trajectories hitting these lines
 slow down and turn back at \em cusps \rm. Since the $(y_1,y_2)$ projection of such trajectories
 \em loiters \rm near the cusp, we call these points \em loitering points \rm and the surfaces
 $\lambda_1,\lambda_2$ we call loitering surfaces.
The distance between the two loitering surfaces is rather small, so most trajectories turn at a sharp U-turn. Both 
loitering surfaces escape to infinity at $O$
and both have coinciding, global minima at $S$: $\lambda_i (1,1,b,c) =  b+\sqrt{b^2+3c}$. Based on this and by using (\ref{VD1}),(\ref{coeffs}) we obtain a global, sufficient condition for the attractivity of the sphere:
\ben
\label{loiterglobal}
\frac{y_3}{z_3} < z_1+z_2+1\,.
\een
In case of spherical objects with radius $y_3=R$ being abraded by spherical abraders with radius $r$ ($z_1=1,z_2=1,z_3=r$) we arrive at the stability result of Bloore (\ref{Bloore2}). In general, (\ref{loiterglobal})
is a global extension of this result (in the frame of the box equations); this condition does not depend on the box ratios $y_1,y_2$ of the abraded object. In addition to the previous observations,
one can also show that trajectories of (\ref{steady}) transversally intersect each loitering surface $\lambda_i$ at most once. Combining these
facts one can globally describe the steady state flow (\ref{steady}). The $(y_1,y_2)$ projection of all trajectories of (\ref{steady}) are either of `Type I': $O\to S$, or of `Type II': $S \to S$,
and they have at most one extremum in either variable, i.e. Type II trajectories are `simple loops' originating and ending at the sphere
(cf. Figure \ref{fig:PDE_LOOP}/a). While the  PDE
 (\ref{Bloore1}) may have more complex structure, 
simple loops appear to be present.
Figure \ref{fig:PDE_LOOP}/b illustrates two trajectories originating at an ellipsoid with axis ratios $y_1=0.83,y_2=0.91$ evolved under (\ref{Bloore1}) and (\ref{steady}), respectively.
While steady state flows correspond to collisional pebble abrasion among well-mixed pebbles and this is an uncommon
geological situation, there exist observations \cite{Nielsen} which strongly suggest that non-monotonic evolution of box ratios may be present in some geological settings.
A detailed description of this phenomenon is provided by Kuenen \cite{Kuenen_Surf}.

 We return to the experimental verification of loops in section 
\ref{s_friction}.
We also remark that Type I. trajectories may correspond to the evolution of pebbles carried by rivers where the abrading environment is the riverbed, represented by a very hard and very large abrader.

\subsection{The self-dual case}\label{s_selfdual}

Now we study the case where two identical pebbles  $\mathbf{y}, \mathbf{z}$ mutually
abrade  each other, 
i.e. we have $\mathbf{y}(t)\equiv\mathbf{z}(t)$ so (\ref{y_box3D})-(\ref{z_box3D}) is again reduced to a 3D system:

\ben
\label{selfdual}
\dot \textbf{y} =  \bar \textbf{F}(\textbf{y},\textbf{y})\,,
\een
which we call the \it self-dual flow \rm with no free parameters and an autonomous direction field in
the ($y_1,y_2$) plane:
\ben \label{selfdual_direction}
y'^C_2=\frac{dy_2}{dy_1}(y_1,y_2)=\frac{F_2^E+2f^b_yF_2^M+f^c_yF_2^G}{F_1^E+2f^b_yF_1^M+f^c_yF_1^G}\,,
\een
where $f^b_y,f^c_y$ are defined in (\ref{coeffs}) by a $\mathbf{z}=\mathbf{y}$ substitution and the
superscript $C$ stands for 'collisional'.
Since (\ref{selfdual}) has time-dependent coefficients, it is difficult to compare it directly to
(\ref{Bloore1}). Nevertheless, it represents the abrasion of pebbles in an environment similar to
itself, i.e. it is the simplest model of the abrasion of segregated pebble populations. The geometry of self-dual flows is simple (visually rather similar to the Gaussian flows illustrated in Figure \ref{fig:components}), in fact
\em all \rm trajectories are of Type I and converge globally to the sphere and in (\ref{selfdual}) we have
\ben \label{selfdual_monoton}
\bar F_i=\dot y_i>0 \qquad i=1,2
\een
on all trajectories, at all times.
 This is not very surprising,
not only because it follows from (\ref{selfdual_direction}) but also because it intuitively matches
(\ref{loiterglobal}). The latter condition was derived for constant $\mathbf{z}$, however, we can
see that it is fulfilled in the self-dual flow on every trajectory, at any time.
Since self-dual flows are the simplest model including segregation, they are closer to the physical process of pebble abrasion on shingle beaches than stationary flows. While self-dual flows have an autonomous 2D component as in (\ref{bracket1}) we showed that such a system might produce homothetic solutions, nevertheless in the collisional box equations there is no such indication.

\section{Friction\label{s_friction} } 

\subsection{Basic assumptions and a semi-local PDE model}

From the previous sections we learnt that \em collisional abrasion, \rm as modeled by
eq. (\ref{Bloore1}) is dominated by the Gaussian and Mean Curvature flows (\ref{components})
thus it converges ultimately to the sphere.
It is not entirely surprising that collisional abrasion is determined by local quantities since
in this model we assume that abrading particles arrive from uniformly distributed directions.
This model is based on the assumption that in the abrasion process the energy level is high enough
to support the free flight of mutually abrading particles.

Frictional abrasion is rather different: here the kinetic energy can be rather low and the orientation
(and thus the global shape) 
of the particle play a key role so we look for a \em semi-local \rm
PDE  model depending both on 
local and on global quantities. Adopting Aristotle's approach,
we regard the radial distance $R$ (measured from the centroid) as our primary variable
but  we also assume frictional abrasion to
depend on the global minimum and maximum of $R$ described by the variables $R_{min}$, $R_{max}$,
respectively:

\ben \label{friction1}
{\p \br(u,v,t)  \over \p t}=-f(R, R_{min}, R_{max}) \bn(u,v,t)\,,\qquad   f>0 \,. 
\een  

Thus our model  may  be regarded as  a generalization of
  that of  Aristotle  (\ref{A1}).
Friction influences abrasion either by sliding or by rolling, and we proceed
by making specific assumptions about these components, starting with sliding.
In shear band experiments
with granular materials it has been observed (\cite{Pena}) that the long axis of elongated particles
gets aligned with the direction of shear. This result indicates that in case of \em sliding, \rm
abrasion will be concentrated on the flat sides of the abraded pebble. This implies that for 
\em sliding friction \rm we can assume

\ben \label{slidingassumption}
\frac{d f(R)}{dR} <  0\,.
\een
In case of \em rolling abrasion \rm the situation is radically different. Rolling is essentially
a 2-dimensional motion, the plane of which is determined by the kinetic energy.
If this energy is so high that the pebble can even roll over the point furthest from the
centroid then we are at the transition to collisional abrasion. There is also
a lower energy threshold below which rolling is not possible: this corresponds
to a trajectory with minimal maximum distance. Rolling occurring between these two
critical energy levels results in a non-monotonic abrasion law, i.e. we assume
that abrasion speed is small for radii to the minimum, zero at the maximum and
larger in-between, i.e we assume the existence of a radius $R_\star$, $R_{min} \leq R_\star \leq R_{max}$ for which we have

\ben \label{rollingassumption}
0 \leq f(R_{min}) \leq f( R_{\star}) \geq f(R_{max}) =  0\,.
\een
The non-monotonicity of the abrasion
law could be explained by observing that surface points corresponding to very large radii are hardly
affected, because they are not part of the typical trajectory.
Points corresponding to very small radii are part of typical trajectories,
however, they will abrade less than points with larger radii along the
same trajectories. In other words: rolling is essentially a planar problem,
the plane of rolling prefers smaller radii because of
energy minimization, however, from among radii in the
plane of rolling, larger radii will be abraded faster (in accordance with Firey's (\cite{Firey}) argument).

The  above ideas can be translated into
what we shall call a semi-local PDE model,
i.e. we propose a definite form for for $f$ in (\ref{friction1}). Here we introduce
separate terms for sliding and rolling with independent coefficients $\nu_s, \nu_r$,
respectively and use the dimensionless ratios $r_1=R/R_{min}, r_2=R/R_{max}$:
\ben \label{friction10}
f(R, R_{min}, R_{max})=\nu_sf_s(r_1,r_2)+\nu_rf_r(r_1,r_2)=\nu_sr_2r_1^{-n}+\nu_rr_2(1-r_2^n)\,.
\een
For sufficiently high values of $n$, this model appears to capture most essential physical 
features of the processes we are aiming to describe. The  PDE  model
is clearly not unique but  provides a simple basis for a qualitative analysis.

\subsection{Unified model for collisional and frictional box flows}

In the box approximations, assumptions (\ref{slidingassumption}) and (\ref{rollingassumption})
translate into

\ben \label{boxassumptions}
\dot u_1 \leq \dot u_2 \leq  \dot u_3 \leq 0, \qquad \dot u_2 \leq \dot u_1 \leq  \dot u_3 \leq 0,
\een
respectively. In the $n \to \infty$ limit the  PDE defined by
 (\ref{friction1}),(\ref{friction10}) yields for the box flows
\ben
\dot u_1  = - \nu_s y_1  -  \nu_r y_1,\quad \dot u_2  = -\nu_r y_2, \quad \dot u_3 = 0\,,
\een
which is equivalent to
\ben 
\label{yy_box3D_friction}
\dot \mathbf{y}  =  \mathbf{F}^f(\mathbf{y},\nu_s,\nu_r) =\frac{1}{y_3}(\nu_s\mathbf{F}^S+\nu_r\mathbf{F}^R)\,,
\een
where
\ben \label{f_components}
\mathbf{F}^S=-\left[ y_1,0,0\right]^T \,, \qquad \mathbf{F}^S=-\left[y_1,y_2,0\right]^T. 
\een
These equations satisfy the conditions (\ref{boxassumptions}).
Similarly to the  self-dual  flows in the collisional model,
we can observe that while the full flow is 3D, both component flows (Sliding, Rolling)
can be reduced to simple planar flows:
\ben\label{friction_direction}
y'^S_{2} =  \frac{F_2^S}{F_1^S} = 0, \qquad \qquad y'^R_{2}  =  \frac{F_2^R}{F_1^R}= \frac{y2}{y1}
\een
and in (\ref{yy_box3D_friction}) we have
\ben \label{friction_monoton}
F^f_i=\dot y_i\leq 0\qquad i=1,2\,.
\een
We can now simply add collisional and frictional flows 
(\ref{y_box3D})-(\ref{z_box3D}), (\ref{yy_box3D_friction})
to obtain the unified equations (denoted by superscript $u$) for the two-body problem:
\bea
\label{yf_box3D}
\dot \mathbf{y} & =&  \mathbf{F}^u(\mathbf{y},\mathbf{z},\nu_s,\nu_r) =\bar \mathbf{F}(\mathbf{y},\mathbf{z})+
\mathbf{F}^f(\mathbf{y},\nu_s,\nu_r) \\
\label{zf_box3D}
\dot \mathbf{z} & = & \mathbf{F}^u(\mathbf{z},\mathbf{u},\nu_s,\nu_r) =\bar \mathbf{F}(\mathbf{z},\mathbf{y})+
\mathbf{F}^f(\mathbf{z},\nu_s,\nu_r) \,.
\eea

\subsection{The steady-state  flow in the presence of friction: experimental evidence of simple loops} 

We study the steady-state flows in the presence of friction, defined by the 3D system:
\ben
\label{yf_stat}
\dot \mathbf{y}  = \mathbf{F}^u(\mathbf{y},\mathbf{z}(0),\nu_s,\nu_r) =\bar \mathbf{F}(\mathbf{y},\mathbf{z}(0))+
\mathbf{F}^f(\mathbf{y},\nu_s,\nu_r)\,. 
\een
The basic geometry is similar to the collisional case, described in subsection
\ref{s_stat}: we can only observe Type I ($O\to S$) and Type II ($S \to S$)
trajectories. To verify the latter we conducted a simple table-top experiment
in the spirit of Kuenen \cite{Kuenen_Gauss}. In each experiment we rolled 5 rounded chalk pieces
of approximately 9.8mm diameter surrounded by approximately 400 hard, plastic balls
of diameter 1.8mm in a horizontal glass cylinder at 5 rpm for approximately 24hrs.
The box dimensions $u_1,u_2,u_3$ were measured every hour.
($u_3$ was the largest diameter, $u_2$ the largest diameter orthogonal
to $u_3$ and $u_1$ was the diameter orthogonal both to $u_3$ and $u_2$.)
The data for the 5 chalk pieces was averaged.
For the computations we used the unified frictional equations (\ref{yf_stat}) and the coefficients
b,c were computed from (\ref{coeffs}). In case of abrading plastic balls with radius $r=0.9$ we had $b=r=0.9, c=r^2=0.81$, respectively.  Initial conditions were set identical to the experiments and the friction
coefficients $\nu_S, \nu_R$ were selected to achieve a good match.

\begin{figure}[!ht]
\begin{center}
\includegraphics[width=80 mm]{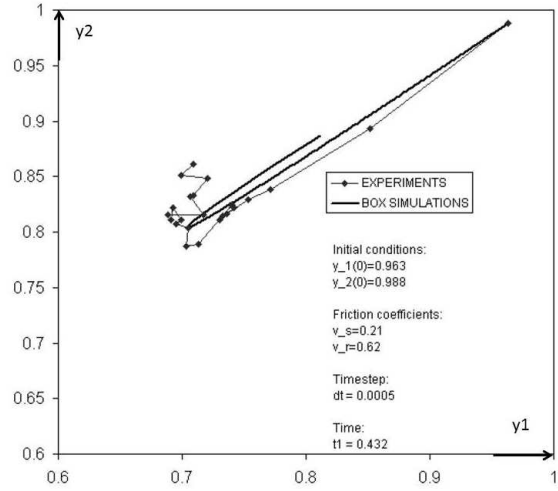}
\end{center}
\caption{The stationary flow in the presence of friction. Observe the marked
'simple loop' both in the numerical and the experimental data}.\label{fig:expts_loop}
\end{figure}

As we can observe in figure \ref{fig:expts_loop}, both  experiments and  numerical simulations show fair quantitative agreement, both displaying the simple loop predicted in subsection \ref{s_stat}. 

\subsection{The self-dual flow in the presence of friction and the existence of stable, nontrivial attractors} \label{s_attractors}

Analogously to the self-dual collisional flows (\ref{selfdual}) 
we define the self-dual frictional flows
as the mutual abrasion of two identical pebbles in the presence of friction:
\ben
\label{yf_self}
\dot \mathbf{y}  =  \mathbf{F}^u(\mathbf{y},\mathbf{y},\nu_s,\nu_r) =\bar \mathbf{F}(\mathbf{y},\mathbf{y})+
\mathbf{F}^f(\mathbf{y},\nu_s,\nu_r)\,. 
\een
which, similarly to (\ref{selfdual_direction}), also
have an autonomous 2D direction field, however, with two free 
parameters $\nu_s,\nu_r$.
We  look for  homothetic solutions
 (\ref{homo}): 
as fixed points of (\ref{yf_self}), i.e. we solve $F^u_1=F^u_2=0$
which is a 2x2 linear system for $\nu_s,\nu_r$, yielding:
\ben
\label{cr}
\nu_s^{cr} = \frac{F_1F_2^S-F_2F_1^S}{F_1^SF_2^R-F_2^SF_1^R}\,, \quad
\nu_r^{cr}  =  \frac{F_2F_1^R-F_1F_2^R}{F_1^SF_2^R-F_2^SF_1^R}\,,
\een
where $F_i$ are from (\ref{yy_box3D}), $F_i^S,F_i^R$ are from (\ref{f_components}). By substituting 
$(y_1=y_{10},y_2=y_{20})$ into (\ref{cr})) we obtain the friction coefficients in the presence
of which (\ref{yf_self}) will have a homothetic solution at $(y_1=y_{10},y_2=y_{20})$.

\begin{figure}[!ht]
\begin{center}
\includegraphics[width=120 mm]{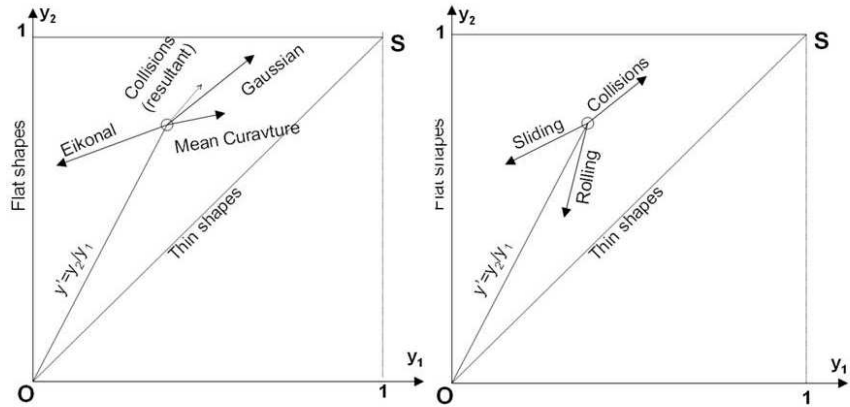}
\end{center}
\caption{'Bracketing' conditions for the existence of homothetic solutions.
a) Equilibrium in the collisional flows. b) Equilibrium in self dual flow in the presence of friction.}
\label{fig:neccond}
\end{figure}
\begin{figure}[!ht]
\begin{center}
\includegraphics[width=100 mm]{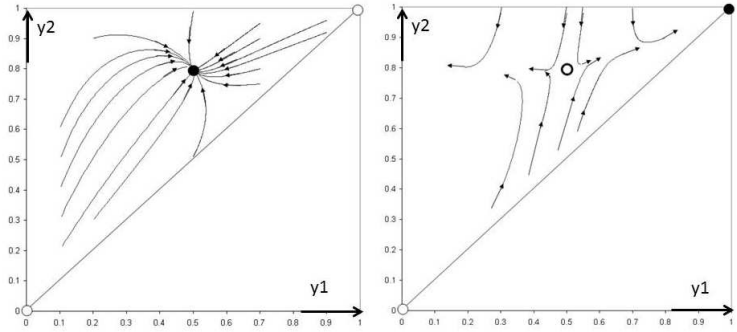}
\end{center}
\caption{\small{The self-dual flow in the presence of friction with nontrivial critical point at $(y_1,y_2)=(0.5,0.8)$. 
a. Friction law according to (\ref{friction10}): single, stable nontrivial attractor, parameter values $(\nu_s=0.622,\nu_r=1.909)$.
b. Friction law according to (\ref{friction_multi}): nontrivial saddle, both the sphere at $(y_1,y_2)=(1,1)$ and infinitely flat shapes at $y_1=0$ are attractive. Parameter values $(\nu_s=1.244,\nu_r=1.909)$.}} 
\label{fig:friction_0508}
\end{figure}

Friction will produce physically observable, stable attractors if both critical parameters
are positive. The condition for this is analogous to the necessary condition (\ref{neccond_collision}) defined
for purely collisional flows. Here we regard the vector sum of the self-dual collisional flow with the frictional
flows. To obtain equilibrium, the former must be 'bracketed' by the latter; this can be expressed based
on (\ref{selfdual_direction}) and (\ref{friction_direction}) as (cf Figure \ref{fig:neccond}/b):
\ben \label{neccond}
 (y'^C_{2}- y'^S_{2})(y'^C_{2}- y'^R_{2})<0\,.
\een
Condition (\ref{neccond}) is satisfied 
by our equations (\ref{yy_box3D_friction})-(\ref{f_components}) for friction, in fact, the stricter condition
\ben \label{neccond1}
0 = y'^S_{2} \leq y'^C_{2} \leq \frac{y_2}{y_1} \leq y'^R_{2}
\een
is also met, and therefore the  proposed PDE
 (\ref{friction10}) is capable 
 of  predicting the existence of 
  nontrivial attractors in the $(y_1,y_2)$ plane.
Since both $F_1,F_2$ appear to be monotonically decreasing in $y_1,y_2$, the fixed points defined
by (\ref{cr}) are stable attractors, illustrated
in Figure \ref{fig:friction_0508}/a.

\section{Stochastic process: multi-body simulations}\label{s_stochastic}

The unified collisional-frictional equations (\ref{yf_box3D})-(\ref{zf_box3D}) describe the mutual abrasion of two individual
pebbles, however, they also define a  \em Markov process \rm where $\mathbf{y},\mathbf{z}$ are random vectors with \em identical \rm
distributions since they represent two random samples of the same pebble population. 
The evolution of this Markov process (and thus the time evolution of
of the pebble size and ratio distributions) is of prime interest since it determines the physical relevance of the stable
attractors identified in subsection \ref{s_attractors}.
While the analytical investigation of the Markov
process is beyond the scope of this paper, direct simulations are relatively straightforward.
We consider $N$ pebbles out of which we randomly draw two with coordinates
$\textbf{y}^0,\textbf{z}^0$ and run equations (\ref{yf_box3D})-(\ref{zf_box3D}) for a short time period
$\Delta t$ on these initial conditions to obtain the
updated vectors $\textbf{y}^1$,$\textbf{z}^{1}$ . In the simplest linear approximation we have the recursive formula
\bea
\label{Nbody1}
\textbf{y}^{i+1} & = & \Delta t \textbf{F}^u(\textbf{y}^i,\textbf{z}^i)\\
\label{Nbody2}
\textbf{z}^{i+1} & = & \Delta t \textbf{F}^u(\textbf{z}^i,\textbf{y}^i)\,.
\eea
Such an iterative step can be regarded as the cumulative, averaged effect of several collisions between
the two selected pebbles.  Apparently, the $N=2, \Delta t \to 0$ case is 
identical to (\ref{yf_box3D})-(\ref{zf_box3D}).
Previously we investigated the behaviour of the deterministic flows in the
special cases of steady state and self-dual flows.
Multi-body simulations allow 
 the numerical study of the statistical stability of the
flows, i.e. one can assess the stability of the above-mentioned special cases.

\subsection{Stochastic, multi-body simulation of collisional flows}

\begin{figure}[!ht]
\begin{center}
\includegraphics[width=80 mm]{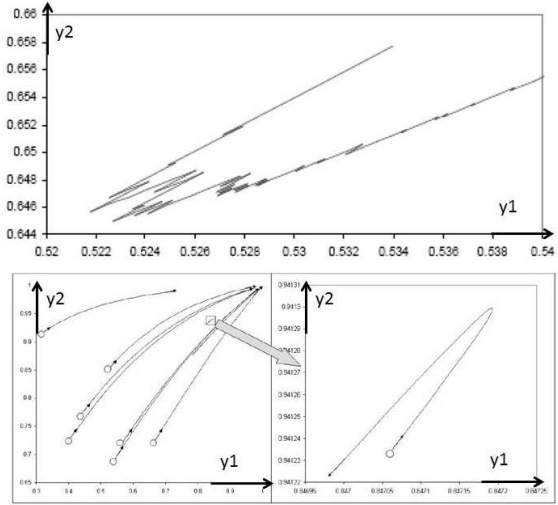}
\end{center}
\caption{Randomized simulation of collisional flows. a) 4-body simulations of the trajectory's 'U' turn in the steady state flow.
b) 9-body simulation of the self-dual flow with random initial conditions.
8 trajectories approach the sphere (as in the deterministic case), one trajectory
turns back (see right hand enlarged version).}\label{fig:multi_collision}
\end{figure}

The characteristic "U"' turn identified in the continuous flow for Type II trajectories
appears to be very robust. In multi-body simulations one can observe the fuzzy
zig-zag geometry of the trajectory near the turning point indicating that 
larger amount of time is spent near the 'loitering' points.
This behaviour is illustrated in Figure \ref{fig:multi_collision}/a for N=4.
The self-dual collisional flow (\ref{selfdual}) proves to be remarkably stable and in multi-body simulations
the trajectories of the continuous flow are often followed rather closely. However, there are
exceptions. Figure \ref{fig:multi_collision}/b illustrates an N=9 body simulation
with random initial conditions, where
8 trajectories converge to sphere (similar to the continuous flow), however, one trajectory
turns back. This deviation from the continuous case can be expected in the vicinity of the sphere
since here the derivatives $\dot y_1, \dot y_2$ converge to zero and so small perturbations
can change their signs.

\subsection{Stochastic, multi-body simulations of the unified collisional and frictional model}\label{s_multibody}

\begin{figure}[!ht]
\begin{center}
\includegraphics[width=80 mm]{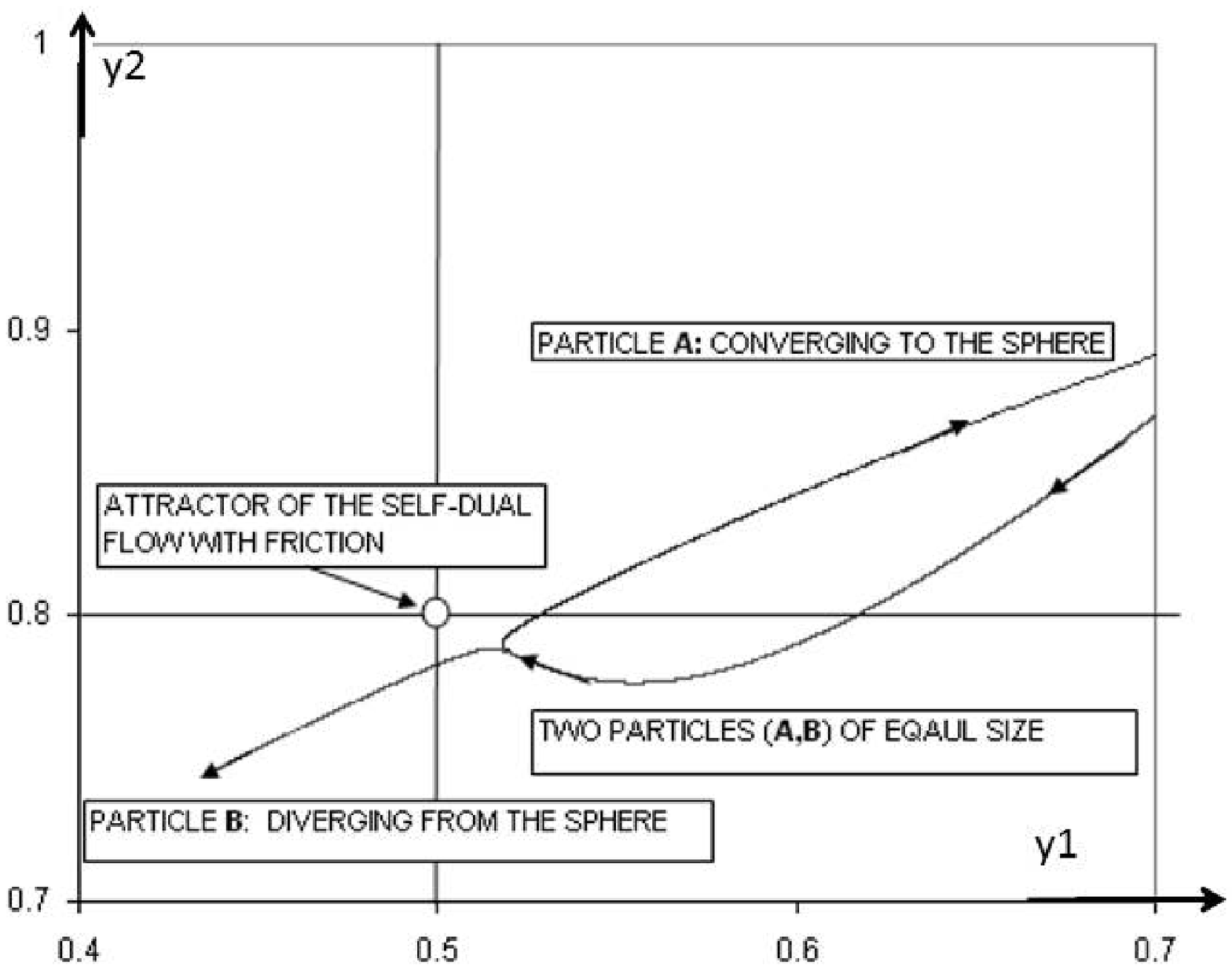}
\end{center}
\caption{Splitting of trajectory in the stochastic simulation (\ref{Nbody1})-(\ref{Nbody2}) of unified collisional and frictional flows (\ref{yf_self}), created by the
mutual abrasion of two almost identical particles. Critical point of the deterministic flow (cf. Figure \ref{fig:friction_0508} ) at
$(y_1,y_2)=(0.5,0.8)$ at parameter values $(\nu_s=0.622,\nu_r=1.909)$).}\label{fig:splitting}
\end{figure}

In subsection \ref{s_attractors} we showed that in the presence of friction stable,
nontrivial attractors appear in the self-dual flows (\ref{yf_self}). While these fix points collectively attract
the abrading objects, the latter mutually repel each other as we illustrate in Figure \ref{fig:splitting},
where two almost-identical pebbles initially follow the same trajectory, however at some
critical time they split. Here we used (\ref{Nbody1})-(\ref{Nbody2}) with $N=2$ and \em almost \rm identical
initial conditions to simulate (\ref{yf_self}).

\begin{figure}[!ht]
\begin{center}
\includegraphics[width=120 mm]{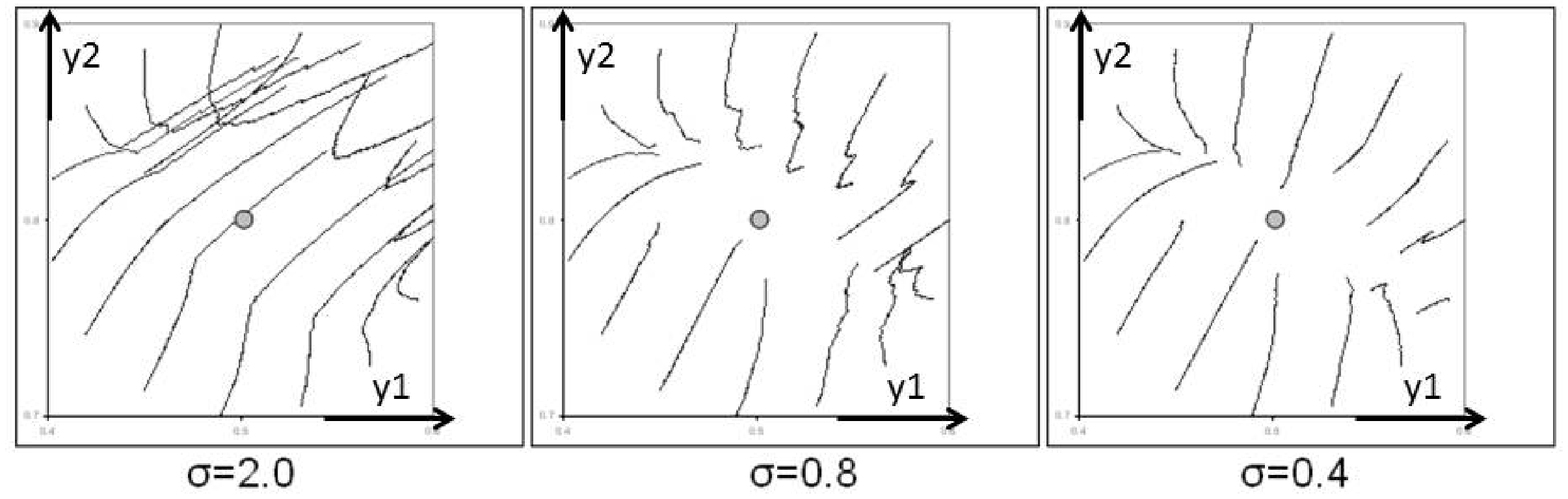}
\end{center}
\caption{The self-dual flow in the presence of friction and segregation.
Critical point of the deterministic flow (cf. Figure \ref{fig:friction_0508} ) at
$(y_1,y_2)=(0.5,0.8)$ at parameter values $(\nu_s=0.622,\nu_r=1.909)$
For weak segregation ($\sigma = 2$) the random trajectories do not converge to the fixed point,
however, with increasing segregation the fixed point of the deterministic flow becomes
attractive also in the stochastic process}.\label{fig:segregation}
\end{figure}

This shows that self dual flows (describing the mutual abrasion of a homogeneous pebble
population consisting of identical
pebbles) are structurally unstable in the sense that the homogeneity
of initial data decays (pebble ratios may mutually repel each other in the abrasion process).
Based on this we expect that in multi-body random simulations according to
(\ref{Nbody1})-(\ref{Nbody2}) the attractors do not appear.
The geological motivation for the study of self-dual flows was the observation
that pebbles are often segregated by size by the wave transport. If this process is
happening on the same (or faster) timescale as abrasion, then the self-dual flows and their
attractors may be stabilized by sustaining the homogeneity
of the initial data, i.e. by selecting only nearly similar pebbles for the collision process.
This could be implemented by
introducing a correlation $r$ between the random vectors $\mathbf{y},\mathbf{z}$;
our observation of the instability corresponds to the uncorrelated $r=0$ case. The fully
correlated  $r=1$ case is essentially identical to the deterministic evolution of (\ref{yf_self}).
Another option to implement segregation (possibly closer to the geological process)
is to omit pebbles from the simulation the maximal size $y_{3,i}$ of which differs from \em all other \rm pebble sizes $y_{3,j}$
by a prescribed factor. For each pebble the coefficient
\ben
\sigma_i = max_j\left|\log\left(\frac{y_{3,i}}{y_{3,j}}\right)\right|
\een
was computed and pebbles with $\sigma_i > \sigma$ were omitted. Increasing segregation
effect corresponds to decreasing values of $\sigma$; this is illustrated in Figure \ref{fig:segregation}.
Observe that for weak segregation $\sigma = 2$ the stable fixed point of the deterministic flow
vanishes while for stronger segregation it persists.

\section{Conclusions}

Our goal was to create a framework in which the interaction of mutual
particle abrasion with global transport and segregation can be studied.

\subsection{Mutual abrasion versus individual abrasion}

As a first step we  
radically simplified Bloore's classical PDE, describing the general
collisional abrasion process of an individual pebble. Our system of ODEs, called the box equations,
attempts to track the evolution of maximal pebble size and main pebble proportions.
While the  box equations are merely a heuristic simplification of the 
PDE describing individual abrasion, they admit the global, analytic
study of {\em mutual abrasion.} To achieve this, 
we incorporated the results of Schneider and Weil \cite{SchneiderWeil},
V\'arkonyi and Domokos \cite{VarkonyiDomokos}, providing the general interpretation of the coefficients in Bloore's PDE.
We arrived at the system of 6 coupled ODE's ,
defining the mutual abrasion
of two pebbles. We investigated two special cases of this rich process 
when  the 6D system
reduces to a 3D system: the steady-state
flows where one pebble is regarded as constant, and the self-dual flows where both
pebbles and abraders are treated  identically. 

\subsection{Special cases: steady state and self-dual flows}

Steady state flows may serve as a skeleton model of abrasion of \em well mixed
\rm pebble populations where the abrading environment has stationary
size distributions. In the box equations we extended  Bloore's local
stability result (\ref{Bloore2}) for nearly-spherical objects abraded by spherical
objects to obtain the sphere's global attractivity criterion. Equation (\ref{loiterglobal})
provides a sufficient, global criterion for the convergence to the sphere,
depending only on the size ratio of the abraded and abrading object
and the box proportions of the latter.  
We found that the steady state box flows admit only two types of trajectories, one of
which appear as \em simple loops \rm  in the plane of the two box
proportions, originating and ending at the sphere. We confirmed the
existence of this 'boomerang effect' 
by direct simulation of the PDE, in
simple table-top experiments and also in stochastic simulations. While
loops may cause some statistical accumulation of  pebble ratios in the
vicinity of the turning point (which we called 'loitering points'),
they certainly do not account for the very clear and striking appearance of dominant
pebble ratios.

We found that the self-dual flows, modeling the abrasion of pebbles in an environment
identical to the abrading pebble, uniformly and globally converge to the sphere. These
flows represent the box equation model for pebble abrasion in segregated environments.

\subsection{Friction}

Since none of the collisional equations accounted for the existence of
homothetic solutions, we added frictional terms, corresponding to
sliding and rolling, with coefficients $\nu_s,\nu_r$, respectively. We
postulated a semi-local  PDE  (\ref{friction10}) based on simple
physical observations and derived box equations describing the
unified, self-dual collisional and frictional process.  We found that
even arbitrarily small amount of friction can stabilize nontrivial
shapes as homothetic solutions in the vicinity of the sphere.  We also
observed that sliding friction stabilizes discoid shapes whereas
rolling friction stabilizes elongated rod-like shapes. At larger values of
the coefficients $\nu_s,\nu_r$, shapes further away from the
sphere can be stabilized. Physically, the increase of the 
coefficients might be due
either to the  increased importance of  frictional efficiency or, more
importantly, to the decrease of the total  energy of the
abrasion process. In the box equations we determined the coefficients
$\nu_s,\nu_r$ corresponding to fixed points with given locations. The
latter data could be extracted from statistical measurements on
shingle beaches and our equations could be used to identify the
associated coefficients. Our  proposed  semi-local  PDE
model is certainly not unique, alternative  PDE's , also
meeting the same criteria, can predict multiple attractors for the
same value of the coefficients;   therefore  several dominant box
ratios could  co-exist  in a pebble environment governed by mutual
abrasion by  collisions and friction. As an example we mention the following model as an alternative to (\ref{friction10}):
\ben \label{friction_multi}
f(R, R_{min}, R_{max})=\nu_s\frac{1-r_2}{r_2}r_1^{-n}+\nu_r r_1\frac{r_1-r_2}{r_1}(1-r_2^n)\,,
\een
which, similarly to (\ref{friction10}), meets the criteria
(\ref{slidingassumption}) and (\ref{rollingassumption}); the corresponding box flow (defined
in the $n\to \infty$ limit) is illustrated in Figure \ref{fig:friction_0508}/b
with a saddle-type critical point.

\subsection{Statistical approach: segregation by transport}

While the existence of attractors in the presence of friction is
promising, the model calls for statistical verification.  Our box
equations in the self-dual  case  describe the mutual abrasion of
two identical pebbles, i.e. the abasion process where
the abraded pebble and abraders are treated completely symmetrically.  
To analyze the statistical stability of the results,
we described the Markov process based on the box equations. 
This process defines the evolution
of pebble size and ratio distributions under the combined
action of collisional and frictional abrasion and segregation.
We ran direct simulations starting
with \em almost identical \rm initial conditions.  In this process,
the colliding pair of pebbles is drawn randomly from the same
distribution. First we modeled abrasion without
transport and segregation. To this end, the draw was uncorrelated 
and we found that the
attractors disappeared: this matched the geological observations
of Carr \cite{Carr}, Landon \cite{Landon} and Bluck \cite{Bluck}
claiming that global transport played an important role in the emergence
of equilibrium shapes.
Next we introduced transport and segregation by size, so
the maximal size of the colliding pairs
was highly correlated and we found that stable attractors were sustained in
the random process. Our model thus predicts that dominant pebble
ratios as stationary equilibrium shapes may robustly appear in abrasion
processes where, in addition to collisions, friction also plays an
important role, and an ongoing global transport process segregates
pebbles based on size. Segregation by shape (also included in our model)
would enhance the stability of these dominant ratios even further.
These predictions match well
geological observations, also, they confirm Aristotle's intuition that the basic
mechanisms of pebble abrasion can only be  understood by including
semi-global and global effects.

\section{Acknowledgements}
The authors thank Andr\'as A. Sipos for his valuable help with the numerical simulation of the
PDE (Figure 2) and and Ott\'o Sebesty\'en for builiding the equipment and conducting the chalk experiments. GD was a Visiting Fellow Commoner at Trinity College, Cambridge
during part of this collaboration.
This research was supported by the Hungarian National Foundation (OTKA) Grant K104601.

\end{document}